# Influence of Introducing High Speed Railways on Intercity Travel Behavior in Vietnam

Van Tho LE[1], Junyi ZHANG[2], Makoto CHIKARAISHI[3] and Akimasa FUJIWARA[4]

[1]Graduate student
E-mail:levanthoutc@yahoo.com
[2]Professor
E-mail: zjy@hiroshima-u.ac.jp
[3]Assistant Professor
E-mail: chikaraishi@hiroshima-u.ac.jp
[4]Professor
E-mail: afujiw@hiroshima-u.ac.jp
Transportation Engineering Laboratory (HiTEL), Graduate School for International Development and Cooperation,
Hiroshima University
(1-5-1 Kagamiyama, Higashi-Hiroshima City, 739-8529, Japan)

**Abstract:**
It is one of hottest topics in Vietnam whether to construct a High Speed Rail (HSR) system or not in near future. To analyze the impacts of introducing the HSR on the intercity travel behavior, this research develops an integrated intercity demand forecasting model to represent trip generation and frequency, destination choice and travel mode choice behavior. For this purpose, a comprehensive questionnaire survey with both Revealed Preference (RP) information (an inter-city trip diary) and Stated Preference (SP) information was conducted in Hanoi in 2011. In the SP part, not only HSR, but also Low Cost Carrier is included in the choice set, together with other existing inter-city travel modes. To make full use of the advantages of each type of data and to overcome their disadvantages, RP and SP data are combined to describe the destination choice and mode choice behavior, while trip generation and frequency are represented by using the RP data. The model estimation results show the inter-relationship between trip generation and frequency, destination choice and travel mode choice, and confirm that those components should not dealt with separately.

## 1. INTRODUCTION

In recent 10 years, the GDP of Vietnam has increasing rapidly, between 5.5 and 8.5 percent annually[1]. The most developed areas are located in Hong Delta River in the north and Mekong Delta River in the south of Vietnam. Due to the nation's topography of narrow and long shape as well as hilly geography, there are some difficulties in performance of the intercity transportation to connecting developed areas via north south corridor in such development situation.

To travelling from Hanoi to Hochiminh City for the distance of around 1700 km with the current intercity transportation situation, it takes about 30 hours by conventional rail and intercity bus, while travel time by airlines is almost 2 hours. The frequency per day is eight and 10 for the two former, and about 16 for the latter. During the holiday, there is a possibility for increasing of frequency and capacity.

The demand for intercity transportation in Vietnam has increasing year by year as a result of economic development and rising population. Moreover, the traffic congestion has occurred in many routes, bus terminals, railway stations and airports, especially in vacation time, with higher frequency and longer congestion time than that of the past. In addition, traffic safety has also become more and more serious with about 12 thousand of fatal people in 2011[2]. Traffic congestion and traffic safety will derive negative influences on regional economic development, national productivity and competiveness, and environmental quality.

To solving the current problem as well as preparing infrastructure to meet the future demand, the Vietnamese government has planning to upgrading the current conventional rail route infrastructure as well as expanding and constructing the express highway and airports. Providing the new ground transportation with high capacity, economically, safety, and environmental friendly is also mentioned in the proposal planning.

To achieving the goals, recently, the Vietnamese government has been considering constructing new HSR line in near future to connecting two biggest cities in Vietnam, namely Hanoi in the north and Hochiminh City in the south. The former has population of more than 6.5 million, whereas that of the later is about 7.5 million by 2010[1]. This line serves for almost all city centers located on costal area and accommodates about 60 percent of the nation's population. If this project will be approved, it will expected to have strong potential influences on economics, politics, land use, and travel behavior in North - South corridor of Vietnam. In addition, some companies recently have operated LCC in some domestic lines. Therefore, it is necessary to conducting researches on the improvement of intercity transportation services to pre-identifying and evaluating alternative proposals. These studies are extremely important for policy makers to establishing the policy for transportation. Ridership of HSR and LCC will be sensitively influenced on these policies as there are not many differences in lever of services between them. The advantages of using HSR are stations located near city center and cheaper in travel cost by comparing with those of LCC. Even the time for access to airport is longer due to its location in suburban area, on-vehicle travel time of LCC is shorter than those of HSR. Thus, the government needs to set up the desirable equilibrium of mode share in transportation market by formulating conscious policy system.

It is obvious that HSR system has several advantages, such as high capacity, safety, fast, and environmental friendly. With its high capacity and frequency, this system is expected to handle part of heavy travel demand in the future via this route.

The HSR system is planned to construct with double track for the distance of 1570 km from Hanoi to Hochiminh City. The designed speed is from 300 to 350 km/h, while estimated ridership is 140 million passengers per year. It takes about six to seven hours to travel to final station. In plan, some sections will be opened in 2020, and in 2035 for whole route[3].

This project meets several constraints, such as limitation of government budget, low technology, and lack of qualify labor force. This mega project costs about 56 billion USD which occupies about 50% of Vietnam GDP 2011 (112 billion[1]). Even the capital will be divided to implementing for long time; investment for each phrase is also enormous. In addition, HSR technology will be depended on foreigner suppliers, since Vietnam has poor technique for producing rolling stock and constructing infrastructure. Therefore, accurate demand forecasting is a requirement for evaluating the feasibility of this project.

As the best knowledge of author, there is not any study at the micro level about the effect of introducing HSR in such context of Vietnam. Hence, the comprehensive questionnaires were designed to conduct a survey in order to achieve several objectives. This paper will reveal the influence of HSR on the vehicle mile travel as well as the trip frequency by introducing the inter-grated intercity travel demand forecasting model to determining influencing factors, such as level of service of travel modes and social demographic characteristics on the decision of mode choice, destination choice, and trip generation.

## 2. LITERATURE REVIEW

Since the first introduction of Shinkansen in Japan in 1964, there are about 10 countries all over the world which have constructed HSR system. With its advantages in journey travel time, HSR is expected to induce additional travel and modal shift. It is reported in King project (1996) that the HSR system in Japan and France have produced as high as 35% for the former and about 30% for the later (as mention in Yao & Morikawa[5]). It is believed that the opening of HSR will endure travel demand, both in short run effects (e.g., route switches, mode switches, changes of destination, and new trip generation) and long term effects (e.g., change in household auto ownership and spatial real- location of activities). For the Tokyo-Nagoya-Osaka corridor (about 500km), the induced travel accounts for 16.5% and 14.5% of travel demand which measured in vehicle mile travel and number of trip respectively[5]. In addition, the demand for HSR will be potential affected by the quality of service of other modes, especially, if the Low Cost Carries (LCC) (Low Cost Airlines) is available.

Changing of destination choice in non-business trip is mainly based on tourism activity due to the location of visiting friends or relatives are fixed, while there are varieties of places for the travelling purpose. The tourism activity is affected by many factors, such as the tourism resources and facility, the motivation of travelers as well as their characteristics, and the accessibility to the destination. The first two former factors are difficult to manage, but the later can be controlled by transportation policy makers as they can setup or modify level of service of transportation modes. In the context of transportation market including HSR and LCC, there is no research on integrated tourism travel demand model for the best knowledge of author. However, there are some studies dealing with tourism behavior which direct or indirect refer to the available of HSR. The research of Masson and Petiot[6] stated the possible effect of HSR between Perpignana (France) and Barcelona (Spain) on reinforcement of tourism attractiveness by using core-periphery model. The HSR could reinforce the agglomeration of the tourism industry on Barcelona which is more developed area than Perpignana. This study concerns about the tourism destination development, but not modeling for tourism travel demand. In addition, the study of Albalate and Bel[7] referenced on HSR in five countries, namely Japan, France, Germany, Spain, and Italy, revealed that tourism and service sector are the activities favored from the construction of HSR.

Some researchers prove that HSR is the dominant mode in the distance less than 800 km, and that of airlines is more than 1.000 km. However, in those studies, the transportation market is without LCC. In addition, the HSR has several advantages, such as stations are located in the urban center, hence, easy to access and egress; very rare cancel and delay; high punctuality. Whereas the LCC fare is cheaper than that of the HSR, but the chance for booking is hard in the meantime of scheduling, and higher rate for cancel and delay. When the different in travel time between HSR and airlines is not significant, it can be supposed that the mode share will highly depend on government strategies, such as policies for level of service and development priority for which kind of mode.

In addition, the study on HSR in Japan, France, Germany, Spain, and Italy of Albalate and Bel[7] found that "modal distribution of traffic is affected when HSR start operation with greatest impact on the airline industry". The study of Park and Ha[8] suggests that the Paris-Lyon route (450km), the air transport mode share decreased half, from 30% to 15%. However, with the longer distance like the Paris-Marseilles route (700 km), the share of airline dropped from 45–55% to 35–45% and the Paris-Nice route (900 km) dropped from 55–65% to 50–60%. This implies that the competition of air transport and HSR is highly influenced by travel distance. The more significant decrease of airline modal share occurs for the short and middle trip distance than that of for long trip distance. As the author's best knowledge, the scientific background of the research about the relationships between trip components in the situation that both HSR and LCC are operated has not been conducted yet. Therefore, the aims of this research focus on exploring the impacts of trip characteristics on mode choice, destination choice behavior, and trip generation in the context of transportation market including LCC.

## 3. INTEGRATED INTERCITY TRAVEL DEMAND MODEL

In the context of modeling intercity travel, Koppelman[4] and Yao and Morikawa[5] indicated intercity trip components, such as trip generation including trip frequency and purpose, destination choice, mode choice and route choice, are inter-related choices and should be combined in a hierarchical structure to explore the travel behavior.

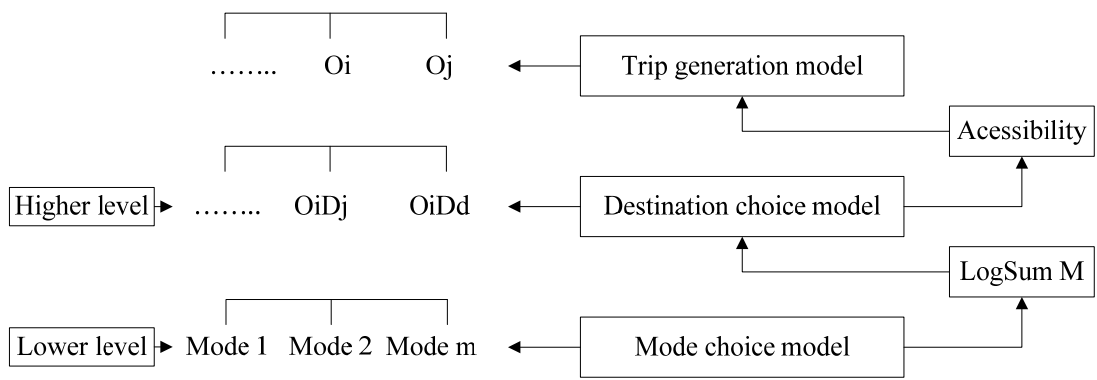

**Figure 1** The nested structure of intercity travel demand model
(based on Yao and Morikawa[5])

This structure can be represented by the nested type structure where the lower nest level affects the upper nest level. The nested structure of intercity travel demand model developed in this study is shown in Figure 1. The proposed methodology includes trip generation and frequency model, destination choice model, and mode choice model. The feedback mechanism represents for the inter-relation between travel choices.

**(1) Mode choice and destination choice models**

This study first develops a modeling system which simultaneous deals with mode choice and destination choice for business and non-business trips, respectively. To represent these two choices, we develop a RP/SP combined nested logit model. The advantages of employing RP/SP approach are to validity issue with SP data and to improve the accuracy of parameters estimate.

Considering the heterogeneity property, the logsum parameter (nest scale parameter) is separately estimated for each individual. The function lets logsum parameter has value between zero and one.

The superscript RP or SP indicates the data type.

a. <u>The business trip model</u>
*RP destination choice model:*
The utility function:
$$U_d^{RP} = V_d^{RP} + \varepsilon_d^{RP} \quad (1)$$
where $V_d^{RP}$ represents the observable components of the utility function of destination $d$, and $\varepsilon_d^{RP}$ is the error term of destination $d$ for RP data.

The observable component of the utility for destination choice $V_d^{RP}$ is specified as:
$$V_d^{RP} = \lambda' C_d^{RP} + \theta_d^{RP} \Gamma_d^{RP} \quad (2)$$
where $C_d^{RP}$ is the explain variable, $\theta_d^{RP}$ represents for logsum parameter associated with nest of destination $d$, and $\Gamma_d^{RP}$ logsum variable (or inclusive value) associated with nest of travel mode choice:
$$\theta_d^{RP} = \frac{\exp(\omega k^{RP})}{1 + \exp(\omega k^{RP})} \quad (3)$$

$$\Gamma_d^{RP} = \log \sum_{m \in M_d^{RP}} \exp \frac{V_m^{RP}}{\theta_d^{RP}} \quad (4)$$

where $k^{RP}$ is the explain variable which is the individual characteristics.

*RP, SP mode choice models:*
The utility function:
$$U_m^{RP} = V_m^{RP} + \varepsilon_m^{RP} \quad (5)$$
(mode m $\in M^{RP}$)
$$U_m^{SP} = V_m^{SP} + \varepsilon_m^{SP} \quad (6)$$
(mode m $\in M^{SP}$)
where $V_m^{RP}$, $V_m^{SP}$ represent the observable components of the utility function of travel mode $m$, and $\varepsilon_m^{RP}$, $\varepsilon_m^{SP}$ are the error terms of travel mode $m$ for RP and SP data.

The observable components of the utility for travel mode choice $V_d^{RP}$, $V_m^{SP}$ are specified as:
$$V_m^{RP} = \beta' X_m^{RP} + \alpha' W_m^{RP} \quad (7)$$
$$V_m^{SP} = \frac{\mu}{\theta_d^{RP}} (\beta' X_m^{SP} + \gamma' Z_m^{SP}) \quad (8)$$
where $X_m^{RP}$, $X_m^{SP}$ are the common attribute vectors; and $W_m^{RP}$, $Z_m^{SP}$ are the specific attribute vectors for the RP and SP data.

The relationship of variations between RP and SP model:
$$\text{Var}(\varepsilon^{RP}) = \frac{\mu^2}{\theta_d^{RP2}} \text{Var}(\varepsilon^{SP}) \quad (9)$$

The joint probability of an individual's choice can be described as:
$$P_{nm}^{RP}(d) = P_{nd}^{RP} P_{nm|d}^{RP} \quad (10)$$
where $P_{nd}^{RP}$ is the marginal destination $d$ probability, $P_{nm|d}^{RP}$ is the conditional probability of travel mode m being chosen given destination $d$ being chosen.

The marginal and conditional probabilities can be derived as:
$$P_{nd}^{RP} = \frac{exp(V_d^{RP})}{\sum_{d' \in D^{RP}} exp(V_{d'}^{RP})} \quad (11)$$

$$P_{nm|d}^{RP} = \frac{\exp\left(\frac{V_m^{RP}}{\theta_d^{RP}}\right)}{\sum_{m' \in M^{RP}} \exp\left(\frac{V_m^{RP}}{\theta_d^{RP}}\right)} \quad (12)$$

The probability of choosing mode m in SP model:
$$P_{nm}^{SP} = \frac{\exp(V_m^{SP})}{\sum_{m' \in M^{SP}} \exp(V_{m'}^{SP})} \quad (13)$$

The log-likelihood function of the RP, SP combined model is:
$$\ln L^{RP+SP} = \ln L^{RP} + \ln L^{SP} \quad (14)$$
where $LL^{RP}$, $LL^{SP}$ are the log-likelihood functions for RP and SP data set.
$$\ln L^{RP} = \sum_{n=1}^{N^{RP}} \sum_{d=1}^{D^{RP}} \sum_{m=1}^{M^{RP}} \ln(P_{nm|d}^{RP})^{\delta_m} (P_{nd}^{RP})^{\delta_d} \quad (15)$$
where $\delta_m$, $\delta_d$ are the dummy variable for individual n choosing mode $m$ travel to destination $d$ in RP model.
$$\ln L^{SP} = \sum_{n=1}^{N^{SP}} \sum_{m=1}^{M^{SP}} \ln(P_{nm}^{SP})^{\delta_m} \quad (16)$$
where $\delta_m$ is the dummy variable for individual n choosing mode $m$ in SP model.

b. <u>The non-business trip model</u>
*RP, SP destination choice model:*
The utility functions:
$$U_d^{RP} = V_d^{RP} + \varepsilon_d^{RP} \quad (17)$$
$$U_d^{SP} = V_d^{SP} + \varepsilon_d^{SP} \quad (18)$$

where $V_d^{RP}$, $V_d^{SP}$ represent the observable components of the utility function of destination $d$, and $\varepsilon_d^{RP}$, $\varepsilon_d^{SP}$ is the error term of destination $d$ for RP and SP data.

The observable component of the utility for destination choice $V_d^{RP}$, $V_d^{SP}$ are specified as:

$$V_d^{RP} = \lambda' C_d^{RP} + \theta_d^{RP} \Gamma_d^{RP} \quad (19)$$

where $C_d^{RP}$ is the explain variable, $\theta_d^{RP}$ represents for logsum parameter associated with nest of destination $d$, and $\Gamma_d^{RP}$ logsum variable (or inclusive value) associated with nest of travel mode choice:

$$\Gamma_d^{RP} = \log \sum_{m \in M_d^{RP}} \exp \frac{V_m^{RP}}{\theta_d^{RP}} \quad (20)$$

$$\theta_d^{RP} = \frac{\exp(\omega k^{RP})}{1 + \exp(\omega k^{RP})} \quad (21)$$

and $V_d^{SP} = \varphi' Q_d^{SP} + \theta_d^{SP} \Gamma_d^{SP} \quad (22)$

where $Q_d^{SP}$ is the explain variables, $\theta_d^{SP}$ represents for logsum parameter associated with nest of destination $d$, and $\Gamma_d^{SP}$ logsum variable (or inclusive value) associated with nest of travel mode choice:

$$\Gamma_d^{SP} = \log \sum_{m \in M_d^{SP}} \exp \frac{V_m^{SP}}{\theta_d^{SP}} \quad (23)$$

$$\theta_d^{SP} = \frac{\exp(\omega k^{SP})}{1 + \exp(\omega k^{SP})} \quad (24)$$

*RP, SP mode choice models:*
The utility functions:
$$U_m^{RP} = V_m^{RP} + \varepsilon_m^{RP} \quad (25)$$
(mode m $\in M^{RP}$)
$$U_m^{SP} = V_m^{SP} + \varepsilon_m^{SP} \quad (26)$$
(mode m $\in M^{SP}$)

where $V_m^{RP}$, $V_m^{SP}$ represent the observable components of the utility function of travel mode $m$, and $\varepsilon_m^{RP}$, $\varepsilon_m^{SP}$ are the error terms of travel mode $m$ for RP and SP data.

The observable component of the utility for travel mode choice $V_d^{RP}$, $V_m^{SP}$ are specified as:

$$V_m^{RP} = \beta' X_m^{RP} + \alpha' W_m^{RP} \quad (27)$$

$$V_m^{SP} = \frac{\mu}{\theta_d^{RP}} (\beta' X_m^{SP} + \gamma' Z_m^{SP}) \quad (28)$$

where $X_m^{RP}$, $X_m^{SP}$ are the common attribute vectors; and $W_m^{RP}$, $Z_m^{SP}$ are the specific attribute vectors for the RP and SP data.

The relationship between RP and SP model:

$$\text{Var}(\varepsilon^{RP}) = \frac{\mu^2}{\theta_d^{RP^2}} \text{Var}(\varepsilon^{SP}) \quad (29)$$

The joint probability of an individual's choice can be described as:

$$P_{nm}^{RP}(d) = P_{nm|d}^{RP} P_{nd}^{RP} \quad (30)$$

where $P_{nd}^{RP}$ is the marginal destination $d$ probability, $P_{nm|d}^{RP}$ is the conditional probability of travel mode $m$ being chosen given destination $d$ being chosen.

The marginal and conditional probabilities can be derived as:

$$P_{nd}^{RP} = \frac{exp(V_d^{RP})}{\sum_{d' \in D^{RP}} exp(V_{d'}^{RP})} \quad (31)$$

$$P_{nm|d}^{RP} = \frac{\exp\left(\frac{V_m^{RP}}{\theta_d^{RP}}\right)}{\sum_{m' \in M^{RP}} \exp\left(\frac{V_m^{RP}}{\theta_d^{RP}}\right)} \quad (32)$$

The joint probability of an individual's choice can be described as:

$$P_{nm}^{SP}(d) = P_{nm|d}^{SP} P_{nd}^{SP} \quad (33)$$

where $P_{nd}^{SP}$ is the marginal destination $d$ probability, $P_{nm|d}^{SP}$ is the conditional probability of travel mode $m$ being chosen given destination $d$ being chosen.

The marginal and conditional probabilities can be express as:

$$P_{nd}^{SP} = \frac{exp(V_d^{SP})}{\sum_{d' \in D^{SP}} exp(V_{d'}^{SP})} \quad (34)$$

$$P_{nm|d}^{SP} = \frac{\exp\left(\frac{V_m^{SP}}{\theta_d^{SP}}\right)}{\sum_{m' \in M^{SP}} \exp\left(\frac{V_m^{SP}}{\theta_d^{SP}}\right)} \quad (35)$$

The log-likelihood function of the RP, SP combined model is:

$$\ln L^{RP+SP} = \ln L^{RP} + \ln L^{SP} \quad (36)$$

where $\ln L^{RP}$, $\ln L^{SP}$ are the log-likelihood functions for RP and SP data set.

$$\ln L^{RP} = \sum_{n=1}^{N^{RP}} \sum_{d=1}^{D^{RP}} \sum_{m=1}^{M^{RP}} \ln (P_{nm|d}^{RP})^{\delta_m} (P_{nd}^{RP})^{\delta_d} \quad (37)$$

$$\ln L^{SP} = \sum_{n=1}^{N^{SP}} \sum_{d=1}^{D^{SP}} \sum_{m=1}^{M^{SP}} \ln (P_{nm|d}^{SP})^{\delta_m} (P_{nd}^{SP})^{\delta_d} \quad (38)$$

where $\delta_m$, $\delta_d$ are the dummy variable for individual n choosing mode m travel to destination d in RP, SP model.

**(2) Trip generation models**

The regression models are employed to estimate for the trip generation with trip frequency as dependent variable, and individual characteristics and accessibility as explain variables. The regression is used for non-business trip generation while negative binomial regression is used for modeling count variables to capture over-dispersed count outcome variables of business trip production.

Let $x = \sum \delta_l G_l \quad (39)$

where $\delta_l$ are the parameter of $G_l$ which are the explain variables including accessibility and personal characteristics.

Then the regression model is:
$$Y = x \quad (40)$$

Table 1 Summary of survey data

| RP/SP data | | | | |
|---|---|---|---|---|
| Survey method | Interview | | | |
| Survey time | Nov 2011 | | | |
| Survey place | Hanoi | | | |
| | | Business | | Non-business |
| | | Mode choice | Destination choice | Mode choice & Destination choice |
| Number of trip | RP | 407 | 407 | 446 |
| | SP | 2432(608) | | 1216(608) |
| | | Destination choice | | Destination choice |
| Number of individual | | 247 | | 524 |

*Note:* The numbers of respondents are given in parentheses

The negative binomial regression model is:
$$Y = 1/(1+1/\exp(x)) \quad (41)$$
It is well-known from literature as well as practical with the use of accessibility measure for the expected maximum utility of individual from origin region i to destination region j

$$\text{Accessibility}_i = \frac{1}{\mu_3} \ln \left( \sum_D \exp(\mu_3 V_{ij}) \right) \quad (42)$$

where $\mu_3$ is the inclusive variable of destination choice level, $V_{ij}$ is the utility function from the origin region i to destination region j, and $D$ is the destination alternatives set for origin i.

The proposed nested structure provides the ability to capture the influence of travel conditions on the trip generation via the measurement of accessibility.

## 4. DATA

For this purpose, a comprehensive questionnaire with both Revealed Preference (RP) and Stated Preference (SP) questions was designed for conducting a survey. This survey aims to first investigate the people's current domestic inter-city travel behavior and then to capture the future travel mode choice and destination choice behavior under the assumption that the HSR will be introduced along North - South corridor of Vietnam. The questionnaire includes two main parts. In the RP section, respondents were asked to fill in trip diary by all intercity trip generation in the past one year, whereas SP part was carefully tailored to examining the choice intention on stated mode choice and destination choice.

The survey was conducted in nine places in Hanoi, Vietnam in November 2011. Survey place covers all of Hanoi area which is included urban-core, urban ring and sub-urban. Respondents were randomly selected in the designed places and interviewers conducted survey by face to face communication. The people who have inter-urban travel in the North-South corridor were selected to answer full set of questionnaire, whereas the others were asked for social demographic characteristics.

In the RP part, respondents were asked to fill in the trip information depend on trip purposes. The data about origin and destination, travel cost, travel time, access and egress time, etc were collected for all trip purpose, while the information about the travel party were used for estimate non-business model. The SP survey was conducted to understand how respondent would reply to different level of services. The interviewer first introduces and describes the HSR to the respondent. Then, the respondent was given a set of questionnaire, in which the hypothetical attributes of each travel mode, such as travel cost, travel time, and service frequency are systematically varied.

The summary of survey data and summary of data characteristics are shown in Table 1, and Appendix A, respectively.

## 5. MODEL ESTIMATION AND RESULTS

The proposed methodology for this research as can be seen in Figure 1 includes trip generation and frequency model, destination choice model, and mode choice model. The feedback mechanism represents for the inter-relation between travel choices.

**(1) Mode choice and destination choice models**

This study develops simultaneous two mode choice and destination choice model systems which

**Table 2** Estimation results of mode choice and destination choice for business trip

| Explanatory variables | | Parameter |
|---|---|---|
| **Destination choice** | | |
| Log of Total GDP (10^6Mil VND) | RP | 8.4697e-01 (***) |
| **Mode choice** | | |
| Travel cost (Mil VND) | All | -1.0421e+00 (***) |
| In-vehicle travel time (h) | All | -3.3467e-02 (***) |
| Access and egress time (h) | All | -7.0339e-02 |
| Constant | | |
|     Bus | RP | -1.3365e+00 (***) |
|     Conventional Rail | RP | -1.6237e+00 (***) |
|     Airlines | RP | 9.6742e-01 (***) |
|     Car | RP | -1.5442e+00 (***) |
|     HSR | SP | 2.0118e+00 (**) |
|     Airlines | SP | 3.3378e+00 (***) |
| Stated dependent | | |
|     Airline | RP | 6.1556e+01 (***) |
|     LCC | RP | 4.1279e+01 (***) |
| Scale parameter for RP and SP data | | 2.7038e-01 (***) |
| **Lambda explanatory variables** | | |
| Occupation* Age (Occupation: Government official, Official staff: 1; otherwise: 0) | RP | -3.2074e-03 |
| Education level* Income (Education: Have university degree or above: 1; otherwise: 0) | RP | -5.3349e-02 (**) |
| Constant | RP | 1.5872e+00 (**) |
| LL0 | | -4043.02 |
| LL1 | | -2914.43 |
| rho | | 0.2791 |
| rho.adj | | 0.2751 |
| VOT (in-vehicle time) | All | 32,114.88 |
| VOT (Access and egress time) | All | 67,497.95 |
| Number of observation | | 928 |

Significant codes:  0 '***' 0.001 '**' 0.01 '*' 0.05 '.' 0.1 ' ' 1

are for business and non-business purpose for accurate representing the decision makers' behavior. The nested structure representing for the influence of mode choice on destination choice with higher level is destination choice, while mode choice is on the lower level.

The estimation results of business trip and non-business trip can be seen in the Table 2 and Table 3 respectively.

The destinations along the proposed HSR route are divided in seven regions as a result of its geography and tourism characteristics. Explain variables include regional characteristics as well as the logsum variable of the maximum utility of mode choice nest.

The available modes in the current situation for any purpose are intercity bus, conventional rail, conventional airlines, LCC, and car. In this study, the airlines and LCC is excluded from the choice set of short distance trip (less than 300 km), while car is not included in the choice set of long distance trip (more than 1300 km) in the modeling system. The choice set of SP survey for business trip of medium and long trip distance includes conventional airlines, LCC and HSR since the assumption of travel cost will be paid by travelers' organizations. For the non-business trip of SP survey, HSR was added to the choice set among all other current available modes.

It is believed that the respondents highly influence by actual choice, and then the dummy variables representing the actual choice in order to capture true state dependent are included in the SP model[1313].

(a) The business trip model estimation results

All the parameters have expected signs and consistent with the literature background.

**Table 3** Estimation results of mode choice and destination choice for non-business trip

| Explanatory variables | | Parameter |
|---|---|---|
| **Destination choice** | | |
| Total number of tourist (Mill) | RP | 0.0240(***) |
| Dummy of Seasonal vacation: Summer: 1, otherwise: 0 | | |
|     destination 2 | RP | -0.2019 |
|     destination 3 | RP | 1.0391 (***) |
|     destination 4 | RP | 1.0152 (***) |
|     destination 5 | RP | -0.6064 (*) |
|     destination 6 | RP | 0.4542(*) |
|     destination 7 | RP | 0.7474 (**) |
| Total evaluation for tourism attraction | SP | 0.3553(**) |
| **Mode choice** | | |
| Travel cost (Mil VND) | All | -0.1378 (***) |
| In vehicle travel time (h) | All | -0.0243 (***) |
| Income | | |
|     Bus user (Mil VND) | All | -0.0505 (***) |
|     Conventional Rail user (Mil VND) | All | -0.0340(**) |
|     Airline user (Mil VND) | All | 0.0134 (.) |
|     Car user (Mil VND) | All | -0.0043 |
|     HSR user (Mil VND) | SP | 0.0051 (**) |
| Constant | | |
|     Bus | RP | 1.1401 (***) |
|     Conventional Rail | RP | 0.9473 (***) |
|     Airlines | RP | 0.9216 (***) |
|     Car | RP | 1.6841 (***) |
|     Bus | SP | 0.6378 (***) |
|     Conventional Rail | SP | 0.5581 (***) |
|     Airlines | SP | 0.2939 (*) |
|     Car | SP | 0.1946 (.) |
|     HSR | SP | 0.2943 (*) |
| Stated dependent | | |
|     Bus | SP | 49.2510 (***) |
|     Conventional Rail | SP | 71.7860 (***) |
|     Airline | SP | 37.7000 (***) |
|     LCC | SP | 74.5810 (***) |
| Scale parameter for RP and SP data | | 4.6683 (***) |
| **Lambda explanatory variables** | | |
| Marital Status*Age (Married: 1; otherwise: 0) | All | -0.0125 (**) |
| Income*Travel party (travel party: with family: 1; otherwise: 0 ) | All | 1.5469 |
| Working status (Working: 1; otherwise: 0) | All | -0.1470 |
| Constant | RP | 1.0696 (**) |
| LL0 | | -5053.84 |
| LL1 | | -3830.42 |
| rho | | 0.2420 |
| rho.adj | | 0.2355 |
| VOT(in-vehicle time) | All | 176,268.4 |
| Number of observation | | 880 |

Significant codes: 0 '***' 0.001 '**' 0.01 '*' 0.05 '.' 0.1 ' ' 1

As in this study, the destination choice is the combination of many provinces/cities, but due to the lack of secondary data, only the total GDP of each region is used as explanatory variable for destination choice level. The estimated coefficient is statistically significant and represent for the influence on the destination choice for business traveler.

Regarding to travel mode choice, the results show that the travel cost, in-vehicle travel time, access and egress time have negative influence on the travel mode choice. All of those parameters are significant except the access and egress time. This may due to the alterative for intercity bus terminal, railway station, and airport are not many for the choice set. The constant terms reflect the inherent preference of travel mode choice (LCC is chosen as base mode). The negative parameter for bus and conventional rail indicate that the businessman has preference to travel by LCC if other variables are the same. However, the opposite situation applies for airline, car and HSR as those parameters have positive signs. All the dummy variables of state dependent have significant coefficients as in the previous studies[13]. This implied that the choices of mode in SP are highly influence of the current behavior.

The government official or official staff with higher age, and the businessman who has university degree or above with lower income are more sensitive for the influence of mode choice on the destination choice of business trip.

The scale parameter for RP and SP data model is less than one which means that the SP data has more noise than that of RP data.

It is intuitive to found that the value of access and egress time is bigger than that of in-vehicle travel time.

   (b) The non-business trip model estimation results

Over all, the model results show indications as expected and consistent with the literature background.

The parameter of total number of tourist and total evaluation for tourist attraction are statistically significant and represent for the influence on the destination choice for traveler. The parameters of dummy for summer vacation as seasonal influence are variety due to the variance of tourist attraction of each region. It is found that tourists are more likely travel to destination 3, 4, 7 where are located of famous beaches. In other words, tourists may be more interested in cooler places as the respondents' location are in Hanoi where is very hot at the summer time. The destination 2 and 5 which are the famous places for the traditional and historical tourism activities have negative influence on travelers' destination choice in summer break time.

For the mode choice model, it is found that the travel cost, and in-vehicle travel time have negative influence on travel mode choice. The negative parameters of income for bus and conventional rail indicate that tourists with same income have preference on LCC if other variables are same. Whereas the travelers with higher income are likely choose airline or HSR for their nonbusiness trip. All constant terms have positive signs indicate that tourists have not preference to travel by LCC if other variables are same. This may due to the intuitive perception about the low quality of LCC. The stated dependent also shows positive influence which means that the travelers tend to repeat their choices behavior with the experience modes.

The person with higher income travel with family, and working people have significant influence of mode choice on destination choice.

The scale parameter for RP and SP data model is bigger than one which means that the SP data has less noise than that of RP data.

It is found that the value of in-vehicle travel time for business trip is much smaller than that of for non-business trip. This contradicts with other current literatures and need study more to confirm since there is not any reference for such kind of study in Vietnam.

  **(2) Trip generation models**

Only resident based respondents are selected for estimating trip generation by regression model. The result can be seen in the Table 4 and Table 5. All most the parameters are statistically significant.

As the accessibility is calculated from equation (42), any change in the level of service can influence on the trip production. For the business trip, the accessibility has positive sign and significant which means that businessman is likely to increase his travel frequency to the region with higher accessibility. For the utility function using to estimate for accessibility of business trip production, the total GDP of region is employed, hence, the more developed areas will attract more business trip. The significant constant means that some other explanatory variables which have influence on the business trip production are not included in this model since the only one explain variable was used in the model estimation. In order to clearly understand the business production behavior patent, it is essential to get a better understanding of the unobserved factor.

The results of non-business trip model shows that the all the explanatory variables other than married status have significant influence on trip generation at 95% level. The working people might have less constrain in financial, therefore, would provide more recreation trip; higher education level can arouse more interest in tourism and allows a better access to information and knowledge of tourism; higher income level can eliminate monetary constraint to participate in tourism; higher age may have more free time to enjoy tourism; and more ease and convenience of accessibility might increase travel frequency. The negative parameter of male

**Table 4** Trip generation model for business trip

| Explanatory variables | Parameter | Std. Error | z value | Pr(>\|z\|) |
|---|---|---|---|---|
| (Intercept) | -11.0280 | 2.3224 | -4.749 | 2.05e-06 (***) |
| Accessibility | 1.6662 | 0.3368 | 4.947 | 7.55e-07 (***) |
| Theta | 0.7115 | | | |
| Std. Err | 0.0995 | | | |
| 2 x log-likelihood | -830.3670 | | | |
| Null deviance | 262.94 | | | |
| res.deviance | 238.26 | | | |
| Number of sample | 247 | | | |

Significant codes: 0 '***' 0.001 '**' 0.01 '*' 0.05 '.' 0.1 ' ' 1

**Table 5** Trip generation model for non-business trip

| Explanatory variables | Parameter | Std. Error | t-stat | P-value |
|---|---|---|---|---|
| (Intercept) | 0 | - | - | - |
| Occupation (working:1; otherwise: 0) | 0.2197 | 0.0713 | 3.0815(**) | 0.0022 |
| Education (having university or above degree: 1; otherwise: 0) | 0.1933 | 0.0665 | 2.9055(**) | 0.0038 |
| Married status (Married:1; otherwise: 0) | 0.1029 | 0.0860 | 1.1962 | 0.2322 |
| Gender (male: 1; female: 0) | -0.1566 | 0.0639 | -2.4493(*) | 0.0146 |
| Age | 0.0063 | 0.0030 | 2.0648(*) | 0.0394 |
| Income | 0.0207 | 0.0068 | 3.0461(**) | 0.0024 |
| Accessibility | 0.0829 | 0.0361 | 2.2998(*) | 0.0219 |
| R square | 0.605072 | | | |
| Adj.R square | 0.598555 | | | |
| Standard error | 0.714335 | | | |
| Number of sample | 524 | | | |

Significant codes: 0 '***' 0.001 '**' 0.01 '*' 0.05 '.' 0.1 ' ' 1

traveler might confront family commitments since the man in Vietnam is considering as head of family, hence, would have lower probability to participate in tourism.

**(3) Induced travel**

As mention in the introduction part, the construction of HSR system will be divided by several phrases, in which, the first segment will take into operation on 2020. To estimate for the future travel demand, the forecasting information about the socioeconomics and the demographic of the analytical year are needed. Due to the lack of that information, this task is left for the future research.

## 6. CONCLUSION

Since the dramatically increasing travel demand along North-South corridor our country is expected in the future, it might be necessary to have a better transportation system to accommodate such future demand. Under such circumstance, the construction of High Speed Rail (HSR) along North-South corridor of Vietnam is under discussion. The proposed HSR is much faster (300km/h), more comfortable, safer, and more punctual, etc., but the construction and operation is costly and as a result, the ticket will be more expensive. In addition, some companies recently have started operating Low Cost Carrier (LCC) recently. In this context, it is necessary to estimate the future travel demand in Vietnam. Therefore, the purpose of this research is to develop an integrated intercity demand forecasting model incorporating trip generation, destination choice and mode choice by business and non-business trip purpose. For this purpose, a comprehensive questionnaire with both Revealed Preference (RP) and Stated Preference (SP) questions was designed for conducting a survey in Hanoi in 2011. In the RP section, respondents were asked to fill in trip diary by all intercity trip generation in the past one year, whereas, SP part was carefully tailored to examining the choice intention on stated mode choice and destination choice. The RP data is used for estimating the trip generation, while the combination of RP and SP data is employed to calculate destination choice and mode choice model. Trip frequency and individual characteristics are taken into account for estimating trip generation, whereas regional's and transportation characteristics are the input variables for destination choice model. As a result, the influence of the introduction of HSR on the changing in the destination choice behavior can be observed. In addition, the SP data is always claimed by its validity and not based on real market behavior, while RP data can be criticized for insufficient variation in explanatory variable, high level of collinearly and inability to treat non-existing alternative. Thus, the combination of RP and SP data will overcome their disadvantages and improve the reliability for estimate the future modal share in the intercity transportation market. Finally, the study successfully established the integrated inter-city travel demand models for business and non-business trip purpose. It is found that travel time, travel cost, access and egress time, but also stated dependent are influence on the choice of travelers both for business and non-business trip. However, the lacking information about regional characteristics drove this study could not be full represented to capturing induce travel. This is left for further study.

**ACKNOWLEDGMENT**
This research was mainly supported by budget of Japanese Government for "Japanese Grant Aid for Human Resource Development Scholarship (JDS)" program, and partly by the Global Environmental Leader (GELs) Education Program of Hiroshima University, Japan.

# APPENDIX A

**Table 6** Summary of data characteristics

| Individual characteristics | Percentage (%) |
|---|---|
| **Age** | |
| 18,19 | 6.03 |
| 20-29 | 37.36 |
| 30-39 | 28.37 |
| 40-49 | 12.07 |
| 50-59 | 10.01 |
| ≥60 | 6.16 |
| **Gender** | |
| Male | 53.15 |
| Female | 46.85 |
| **Marital status** | |
| Single | 35.94 |
| Married | 61.49 |
| Others | 2.57 |
| **Occupation** | |
| Government officer/ Office staff | 37.48 |
| Industrial laborer | 7.83 |
| Merchant | 7.83 |
| House-wife/ Jobless/ Retied | 11.04 |
| Student, pupil | 18.87 |
| Others | 16.94 |
| **Final academic degree** | |
| Senior high school | 14.51 |
| College/Vocational training | 18.61 |
| Bachelor | 56.61 |
| Master / Doctor degree | 5.01 |
| Others | 5.26 |
| **Income information** | |
| ≤ 1,600 | 11.04 |
| 1,601 - 3,000 | 19.38 |
| 3,001 - 5,000 | 21.57 |
| 5,001 - 10,000 | 27.98 |
| 10,001 - 15,000 | 10.53 |
| 15,001 - 20,000 | 6.55 |
| > 20,000 | 2.95 |

# APPENDIX B

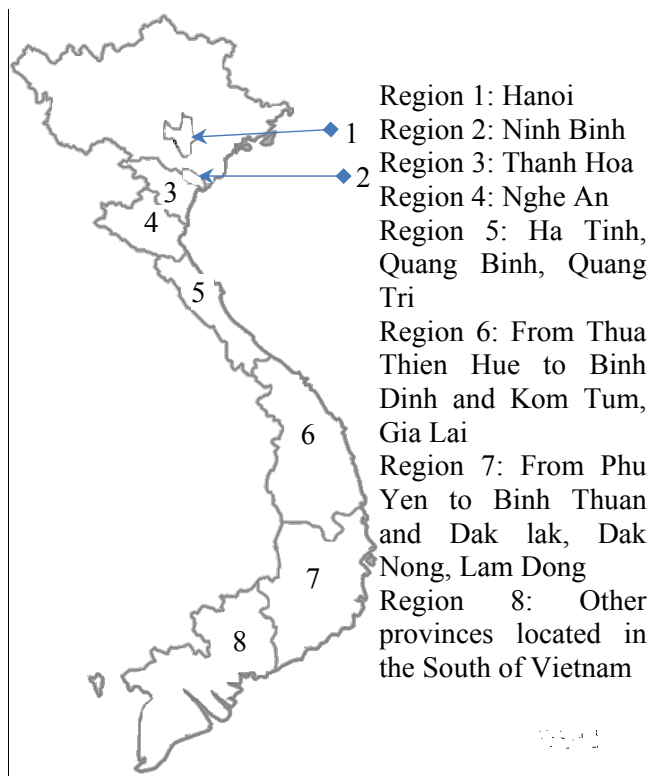

Region 1: Hanoi
Region 2: Ninh Binh
Region 3: Thanh Hoa
Region 4: Nghe An
Region 5: Ha Tinh, Quang Binh, Quang Tri
Region 6: From Thua Thien Hue to Binh Dinh and Kom Tum, Gia Lai
Region 7: From Phu Yen to Binh Thuan and Dak lak, Dak Nong, Lam Dong
Region 8: Other provinces located in the South of Vietnam

**Figure 2 Map of destination alternatives**